\documentclass[twocolumn,aps,showpacs,eqsecnum,10pt]{revtex4} %twocolumn,
\usepackage{latexsym}
\usepackage{amsthm}
\usepackage{amsmath}
\usepackage{amssymb}
\usepackage{epsfig}
\usepackage{color}
\usepackage{units}

\theoremstyle{definition}

\newcommand{\bra}[1]{\langle #1|}
\newcommand{\ket}[1]{| #1 \rangle }
\newcommand{\tr}[1]{{\rm tr}[#1]}

\newcommand{\be}{\begin{eqnarray}}
\newcommand{\ee}{\end{eqnarray}}
\def\idty{{\rm1\mkern -5.4muI}}

\newcommand{\cE}{{\cal E}}
\newcommand{\cI}{{\cal I}}

\newcommand{\cS}{{\cal S}}
\newcommand{\cH}{{\cal H}}
\newcommand{\cA}{{\cal A}}

\newcommand{\ie}{{\it i.e.}}

\newcommand{\0}{|0\rangle}
\newcommand{\1}{|1\rangle}

\newcommand{\Matrix}{{\cal M}}
\newcommand{\Vector}{\vec{v}}

\begin{document}
\graphicspath{{figures/}}

\title{Scrutinizing single-qubit quantum channels: Theory and experiment with trapped ions}
\author{Th.~Hannemann$^{1}$, Chr.~Wunderlich$^{1}$}
\affiliation{$^{1}$Fachbereich Physik, Universit\"at Siegen, 57068 Siegen, Germany}

\author{ Martin Plesch$^{2,4}$, M\'ario Ziman$^{2,3}$, and Vladim\'\i r Bu\v zek$^{2,4}$}
\affiliation{$^{2}$Research Center for Quantum Information, Institute of Physics,  Slovak Academy of Sciences,
       D\'ubravsk\'a cesta 9, 845 11 Bratislava, Slovakia\\
$^{3}$ Faculty of Informatics, Masaryk University, Botanick\'a 68a, 602
00 Brno, Czech Republic \\
$^{4}$QUNIVERSE, L\'{\i}\v{s}\v{c}ie \'{u}dolie 116, 841 04 Bratislava, Slovakia
}

\begin{abstract}
We report experimental implementation of various types of qubit channels using an individual trapped ion.
We analyzed experimental data and we performed tomographic reconstruction of
quantum channels based on these data. Specifically, we studied phase damping channels, where the damping acts either in the $xy$-plane of the Bloch sphere or in an arbitrary plane that includes the origin of the Bloch sphere.
We also experimentally realized and consequently analyzed quantum channels
that in addition to phase damping affect also a polarization rotation.
We used three reconstruction schemes for estimation of quantum channels from experimental data:
{\bf i)} a linear inverse method, {\bf ii)} a maximum likelihood estimation, and {\bf iii)} a
constrained maximum likelihood estimation. We took into account realistic
experimental conditions where imperfect test-state preparations and biased
measurements are incorporated into the estimation schemes. As a result we
found that imperfections present in the process of preparation of test states and as well as in measurements
of the considered ion trap system  do not limit the control of the
implementation of the desired channel. Even imperfect preparation of test state and subsequent measurements still provide sufficient resources for the complete quantum-channel tomography.
\end{abstract}

\pacs{03.65.Wj, 03.67.Lx,}

\maketitle

%%%%%%%%%%%%%%%%%%%%%%%%%%%%%%%%%%%%%%%%%% introduction
%%%%%%%%%%%%%%%%%%%%%%%%%%%%%%%%%%%%%%%%%%%%%%%%%%%%%%%%%%%%%%%%%%%%
\section{Introduction}
%%%%%%%%%%%%%%%%%%%%%%%%%%%%%%%%%%%%%%%%%%%%%%%%%%%%%%%%%%%%%%%%%%%%
Any measurement on a physical system gives a result with limited
accuracy and precision. This property comes from the fact that the
number of data gained by real physical measurements is finite and,
moreover, the measurements are influenced by errors and
imperfections of the experimental setup. Nevertheless, it is the
role of a physicist to interpret measured data and to conclude
about properties of the system as reliably as possible.

Determining the action of an unknown quantum channel by quantum
process estimation enables us to protect and optimally exploit
 quantum systems for communication purposes and also for quantum
computation by designing efficient schemes correcting errors
introduced by the noise (see for instance \cite{nielsen,paris}).
Furthermore, for quantum
computing it is vital to characterize and improve the experimental
performance of quantum logic gates. Quantum process estimation is
needed here to first identify imperfections and decoherence. The knowledge 
of such imperfections then allows us to find remedies for these usually unwanted effects.

Any reconstruction of a quantum channel channel can be viewed as a black-box problem. The black box
(quantum channel) is probed by test states. At the input these states are specifically prepared while at the output
are specifically measured. Based on the correlations between preparations and measurements one can reconstruct 
the action of the quantum channel under consideration.
In this article we investigate how to reconstruct the action of
various engineered quantum qubit channels even in the presence of
imperfect initial state preparation and imperfect detection. We
shall employ two reconstruction schemes which lead to the same
result for ``good'' data. The first of these reconstruction schemes is the inverse method.
In some cases this approach can result
an unphysical estimation of the channel (e.g., the estimated channel is not
completely positive). Small deviations from the physically valid
region are acceptable, because they can be understood as an evidence of
finite statistics. In this case the observed frequencies of specific outcomes of
measurements  are not
perfectly compatible with probabilities as predicted by quantum
theory, but we can use the maximum likelihood method
\cite{fisher,fiurasek,sacchi} in order to
perform the reconstruction. By construction,  the maximum likelihood
method always leads to a physical result. If the violation of the
physical constraints is large, then the data are apparently biased
by some kind of systematic errors, or imperfections in the
experimental setup.

\subsection{Description of qubit channels}
Within the framework of the quantum theory  quantum channels are described by
completely positive trace-preserving linear maps acting on the
system's state space $\cS(\cH)$. For qubits $\cS(\cH)$ consists
of operators $\varrho=\frac{1}{2}(\idty+\vec{r}\cdot\vec{\sigma})$
where $\vec{r}$ is a real vector with $|\vec{r}|<1$,
$\vec{\sigma}=(\sigma_x,\sigma_y,\sigma_z)$ is a vector
of Pauli operators, and $\idty$ represents the identity operator. Therefore, the qubit states can be
illustrated as points inside the so-called Bloch sphere formed
by Bloch vectors $\vec{r}$.
In this {\it Bloch-sphere representation}  quantum channels
$\cE:\cS(\cH)\to\cS(\cH)$ are described as affine transformations
of Bloch vectors, i.e.
\be
\cE:\vec{r}\to M\vec{r}+\vec{v}\,,
\ee
where $M$ is a real $3\times 3$
matrix and $\vec{v}$ is a translation \cite{ruskai}.
In particular, $M_{jk}=\tr{\sigma_j\cE[\sigma_k]}$ and
$v_{j}=\tr{\sigma_j\cE[I]}$. Consequently, 12 real parameters need
to be determined in order to completely specify a qubit channel $\cE$.

According to \cite{ruskai} each qubit channel is unitarily equivalent
to a specific quantum channel with diagonal matrix
$D={\rm diag}\{\lambda_1,\lambda_2,\lambda_3\}$. In particular,
$M=R_1 D R_2$, where $R_1,R_2$ are suitable orthogonal rotations
associated with unitary channels $U_1,U_2$, i.e.
$U_j\varrho U_j^\dagger$ gives Bloch vector transformation
$\vec{r}\to R_j\vec{r}$. The diagonal values
$\{\lambda_1,\lambda_2,\lambda_3\}$ are the so-called singular values of
$M$. If $\vec{v}=\vec{0}$, then the channel is called {\it unital}.
In our analysis we shall be interested in a special subclass
of unital channels - {\it phase damping channels} characterized by the
relation $D={\rm diag}\{\lambda,\lambda,1\}$.

The affine form guarantees that for all matrices $M$ and vectors
$\vec{v}$ the resulting affine map is trace-preserving.
Unfortunately, the analytic conditions of complete positivity are
not simple and illustrative \cite{ruskai}. The complete positivity
is equivalent to positivity of the state
$\omega_\cE=(\cE\otimes\cI)[\Psi_+]$, where $\cI$ stands
for the identity channel and $\Psi_+$ is
a projection onto a vector
$\ket{\psi_+}=\frac{1}{\sqrt{2}}(\ket{00}+\ket{11})$ describing two-qubit maximally entangled state. If a qubit
channel is completely positive, then $|\lambda_x\pm\lambda_y|\leq
|1\pm\lambda_z|$ for all combinations of pluses and minuses. Hence,
for phase damping channels $|\lambda|\leq 1$.

There are many different experimental procedures how to determine
all  channel parameters \cite{paris,chuang,poyatos,dariano,jezek,altepeter}.
In general, such an experiment consists
of three steps:
\begin{description}
\item[i)] preparation of a {\it test state};
\item[ii)] application of an unknown channel;
\item[iii)] measurement of the transformed system.
\end{description}
In principle, the preparation stage can result
in a correlated state of a composite system of the system and some
ancilla. In this case the experimental arrangement is such that
the ancilla is not directly affected by the quantum channel, but the
measurement is performed on the whole system. However, this
procedure is itself an experimental challenge, and from the
experimental point of view, ancilla-free preparations are
preferable.

In what follows we shall consider only ancilla-free experiments. By
measuring the correlations between the channel inputs and outputs we
can determine the channel action and identify the corresponding
affine mapping $\vec{r}\to M\vec{r}+\vec{v}$. The linearity of $\cE$
implies that for the purposes of the complete-channel tomography it
is necessary to use a collection of linearly independent test states
spanning the whole state space. In particular, since for a qubit the
linear span of the state space is four-dimensional it follows that
at least four test states are necessary.

An important assumption of the above model of the experiment is that
the test states are perfectly known, i.e., their preparation is
under complete control. That is experimental imperfections (arbitrarily
small) in preparation of the desired test states must be included
into the description of the test states. If this is the case, then
the channel tomography reduces to state tomography of the output
states.

This paper is organized as follows: in Section II we shall describe
the experimental implementation of the single-qubit channels and
describe the measured experimental data. In Section III we shall
discuss different strategies for complete channel reconstruction. In
Sections IV, V, and VI the tomography of channels using different reconstruction schemes is presented. 
Results and observations are summarized in Section VII.

%%%%%%%%%%%%%%%%%%%%%%%%%%%%%%%%%%%%%%%%%% experiment
%%%%%%%%%%%%%%%%%%%%%%%%%%%%%%%%%%%%%%%%%%%%%%%%%%%%%%%%%%%%%%%%%%%%
\section{Description of the experiment}
%%%%%%%%%%%%%%%%%%%%%%%%%%%%%%%%%%%%%%%%%%%%%%%%%%%%%%%%%%%%%%%%%%%%
In this section the physical realization of various types of quantum
channels is described. These include phase-damping channels acting
in the x-y-plane of the Bloch sphere where the amount of damping is
varied by changing the amplitude of added noise. In addition, phase
damping in an arbitrary basis is exemplified by a quantum channel
where the damping acts in a plane cutting through the Bloch sphere
spanned by the vectors $(0,\sin\pi/4,\cos\pi/4)^T$ and $(1,0,0)^T$.
Furthermore, combinations of phase damping and polarization rotating
channels are implemented with varying rotation angles. A
representative subset of these channels have been chosen for
detailed reconstruction that is presented in Section \ref{sec:III}.

The above mentioned quantum channels have been realized
experimentally using two hyperfine states of the electronic ground
state of an individual electrodynamically trapped $^{171}$Yb$^+$ ion
as a qubit (\mbox{$\0 \equiv |S_{1/2}F=0\rangle$}, \mbox{$\1 \equiv
|S_{1/2}F=1, m_F=0\rangle $})
\cite{Huesmann99,Hannemann02,Wunderlich03,Balzer06} and exposing
this qubit to unitary operations as well as engineered irreversible
dynamics.

In order to take an individual data point that contributes to the
characterization of a quantum channel we proceed as follows:

\begin{description}
\item[i)] The ion is laser cooled.
\item[ii)] The qubit is initialized in the state $\0$.
\item[iii)] The desired
input state for testing the quantum channel is prepared. This is either
one of the eigenstates of $\sigma_k$ ($\sigma_k$ are the Pauli
matrices, $k= x, y, z$) with positive eigenvalue (``spin up''), or the
eigenstate of $\sigma_z$ with negative eigenvalue (``spin down'').
\item[iv)] The qubit is
subjected to the action of the quantum channel.
\item[v)] The qubit is measured in a predetermined basis ($\pm \sigma_j$). Below, in
Section \ref{sec:Measurement}, it is shown how the effect of a
difference in the detection efficiencies for states $\0$ and $\1$,
respectively can be canceled by measuring $\sigma_j$,
but also $-\sigma_j$.
\end{description}
For a given set of parameters characterizing the quantum channel
the sequence i)-v) is then repeated for all combinations of test
states [prepared in the step iii)] and measurements [chosen in the step v)].
Then one parameter of the quantum channel is changed and all
preparation and measurement steps described so far are repeated for
this changed set of parameters. Finally, the complete sequence is
repeated between 100 and 500 times that allows us to extract from these
measurements  relative frequencies $c_{\pm j,\pm k}^{\pm}$. Here
the index $j\in\{x,y,z\}$ indicates that a measurement in $\sigma_j$
direction has taken place when the initially prepared state was
$\varrho_{\pm k}=\frac{1}{2}(I\pm\sigma_k)$, $k \in \{x,y,z\}$. The
superscript $\pm$ indicates the outcome of a measurement $\sigma_j$.
The notation introduced here for the frequencies $c_{\pm j,\pm
k}^{\pm}$ allows for a compact description. Note that not all
possible combinations of test states and measurement directions are
realized experimentally. It is sufficient to prepare four different
test states as described in the step iii) above.

In the remainder of this section the initial preparation of the
qubit before exposing it to the action of the quantum channel, the
measurement in an arbitrary basis, the final read-out, coherent
operations and controlled addition of noise are briefly reviewed.

\subsection{Physical realization of quantum channels}
\subsubsection{The qubit and its initialization}
The experimental setup used here for a quantum-process estimation has
been described elsewhere \cite{Hannemann02,Wunderlich03} and a brief
summary should suffice: The quantum mechanical two-state system used
as a qubit is the S$_{1/2}$ ground-state hyperfine doublet with total
an angular momentum $F= 0,1$ of a single $^{171}$Yb$^+$ ion confined in
a miniature Paul trap (the diameter of 2 mm). The
 \mbox{$\0 \equiv |F=0\rangle  \leftrightarrow  |F=1, m_F=0\rangle \equiv \1$}
transition with Bohr frequency $\omega_0$ is driven by a
quasi-resonant microwave (mw) field with an angular frequency near
$\omega = 2\pi\times 12.6$ GHz. A static magnetic field is applied
to the ion such that the three Zeeman states of the S$_{1/2}$, F=1
manifold are split by about 6 Mhz. The unitary dynamics of the qubit
driven by microwave radiation is virtually free of decoherence, {\it
i.e.} transversal and longitudinal relaxation rates are negligible
\cite{Huesmann99,Hannemann02,Wunderlich03}.

Illuminating the ion with laser light near 369,5 nm, generated by a
frequency-doubled Ti:sapphire laser, exciting the S$_{1/2}$(F=1)
$\leftrightarrow$ P$_{1/2}$(F=0) resonance serves for initial state
preparation in state $\0$ while photon-counting resonance
fluorescence on the S$_{1/2}$(F=1) $\leftrightarrow$ P$_{1/2}$(F=0)
transition driven by light near 369~nm allows for a state-selective
detection of the qubit.
Optically pumping the ion into the metastable $^2D_{3/2}$ level via
the P$_{1/2}$ state is prevented by illumination with light near
935~nm of a diode laser that retrieves the ion to the ground state
via the $|D_{3/2}$, $F$=$1\rangle \rightarrow |[3/2]_{1/2}$,
$F=0\rangle $ excitation \cite{Bell91}. Laser cooling of the ion is
achieved by simultaneously irradiating the ion  for 20~ms with laser
light near 369 nm and 935 nm and mw radiation where the latter is
resonant with the $\0 \leftrightarrow \1$ transition. This is done
before initializing the ion into state $\0$: upon turning the
microwave radiation off the ion is then optically pumped into state
$\0$ with probability $p_0 = 0.92$ in the experiments reported here
\cite{Balzer06}.

\subsubsection{Measurement of the qubit state}
\label{sec:Measurement} To detect the state of the ion it is
irradiated with laser light near \unit[369]{nm}\ for a duration of
\unit[2]{ms}. If the ion is in the S$_{1/2}$(F=1) state, on average
$6.25$ photons are registered during that time interval. Due to
stray light and dark counts on average $0.16$ photons are
registered, if the ion is in the S$_{1/2}$(F=0) state. In both cases
the numbers of registered photons follow a Poissonian distribution.
If more than one photon is registered during the counting interval
the ion is assumed to be in the S$_{1/2}$(F=1) state, and if one
photon or less is registered the state is assumed to be
S$_{1/2}$(F=0). The probability to detect the S$_{1/2}$(F = 1) state
correctly is $\eta_1 = 0.986$ and the probability to detect the
S$_{1/2}$(F=0) state correctly is $\eta_0= 0.989$
\cite{Hannemann02}. This implies that the two outcomes are
associated with positive operators \be
E_0^+&=&\eta_0 \ket{0}\bra{0}+(1-\eta_1)\ket{1}\bra{1}\, ,\\
E_1^+&=&(1-\eta_0) \ket{0}\bra{0}+\eta_1\ket{1}\bra{1}\, .
\ee
The detection efficiencies for both states are not equal, which if
not taken into account adds a slight systematic bias to the
measurement results. This bias is avoided, if for each measurement
in one direction also the measurement in the opposite direction is
taken. This means that when, for example, the $z-$direction is to be
measured, also the direction $-z$ is measured.
Such an ``inverted''
measurement is described by positive operators
\be
E_0^-&=&\eta_0 \ket{1}\bra{1}+(1-\eta_1)\ket{0}\bra{0}\, ,\\
E_1^-&=&(1-\eta_0) \ket{1}\bra{1}+\eta_1\ket{0}\bra{0}\, . \ee
Therefore, combining both measurements into a single one (with equal
probability) and identifying the pairs of outcomes $E_0^+, E_1^-$
and $E_1^+, E_0^-$ will form a new measurement of the $z-$direction
$\Sigma_z$ associated with positive operators
 \be \nonumber
F_+&=&\frac{1}{2}(E_0^++E_1^-)\\
\nonumber &=&
\frac{1}{2}[(\eta_0+\eta_1)\ket{0}\bra{0}+(2-\eta_1-\eta_0)\ket{1}\bra{1}]\\
 &=&\eta\ket{0}\bra{0}+(1-\eta)\ket{1}\bra{1}\, ,
 \ee
 and
 \be
  F_-=\eta\ket{1}\bra{1}+(1-\eta)\ket{0}\bra{0}\, ,
 \ee
where $\eta=(\eta_0+\eta_1)/2$.

As already defined above, we denote by $c^{\pm}_{\pm j,\pm
k}=\tr{E^\pm_{\pm j}\varrho_{\pm k}}$ the frequencies of outcomes
$\pm$ for measurements in the direction $\pm j$ provided that the test
state $\varrho_{\pm k}$ was used. Then, for the frequencies of
measurements $\Sigma_x,\Sigma_y,\Sigma_z$ we obtain
\begin{equation}
\begin{array}{cc}
f_{\pm j,\pm k}=[c^+_{\pm j,\pm k}+(1-c^-_{\mp j,\pm k})]/2
\end{array}
\end{equation}
with $j,k \in \{x, y, z\}$.

A measurement in a given direction is performed in two steps: First,
a unitary transformation of the qubit is performed (Sec.~
\ref{Coherent}) effecting a rotation of the desired measurement axis
onto the z-axis. Second, the qubit is irradiated for \unit[2]{ms}\
with laser light resonant with the S$_{1/2}$(F=1) $\leftrightarrow$
P$_{1/2}$ transition and scattered photons are detected, if the state
$\1$ is occupied.

In summary, collecting  data in the described way
results in the implementation of the measurements $\Sigma_x,\Sigma_y,\Sigma_z$
described by POVM elements
\be
F_{\pm j}=\eta P_{\pm j}+(1-\eta)P_{\mp j}=\frac{1}{2}[I\pm (2\eta-1)\sigma_j]\,,
\ee
where we used the fact that $P_{\pm j}=\frac{1}{2}(I\pm \sigma_j)$
are the projectors onto the eigenvectors of $\sigma_j$ associated
with eigenvalues $\pm 1$.

\subsubsection{Coherent operations}
\label{Coherent}
 Coherent operations on the qubit are achieved by
applying near-resonant microwave radiation with the angular frequency
$\omega$ driving the magnetic dipole transition between the states $\0$
and $\1$. In the reference frame rotating with $\omega$, after
applying the rotating wave approximation, the time evolution
operator determining the evolution of the qubit exposed to linearly
polarized mw radiation reads
 $ U(t)= \exp\left[-\frac{i}{2}t\left( \delta
          \sigma_{\text{z}} + \Omega\sigma_{\text{x}}\right)\right]
 $.
Here, the Rabi frequency is denoted by $\Omega$ and $\delta \equiv
\omega_0-\omega$ is the detuning between the qubit and the applied
radiation. A desired state characterized by a nutation angle $\theta$
and an azimuthal angle $\phi$ on the Bloch sphere is prepared by
driving the qubit initially prepared in the state $\0$ with mw pulses
with appropriately chosen detuning $\delta$, the Rabi frequency
$\Omega$, and the duration $t_{\text{mw}}=\theta/\Omega$, and by
allowing for a free precession for a prescribed time
$t_{\text{p}}=\phi/\delta$: Since the applied mw radiation is
slightly detuned from the qubit resonance near 12.6 Ghz, the qubit
acquires a phase $\phi$ relative to the driving field. Note that
$\phi=0$ when waiting integer multiples of the time
$T_P=2\pi/\delta$ needed for a full precession of the Bloch vector
in the $x-y$-plane.

Typical values in these experiments were $\Omega=\unit[3.25\times
2\pi]{kHz}$ and $\delta=\unit[104.5\times 2\pi]{Hz}$. These
parameters were determined by recording Rabi oscillations over 2--4
periods and by performing a Ramsey-type experiment with mw pulses
separated in time. Also, a Ramsey experiment served to establish the
coherence time of the qubit (\ie, the lifetime of the off-diagonal
elements of the density matrix $\rho$ describing the qubit, often
termed $T_2$ time) that was found to be well over one second
\cite{Wunderlich03}.

\subsubsection{Phase damping}
 \label{sec:phase_damping}
Magnetic field noise may be applied during a prescribed time to the
qubit in order to induce phase damping, that is, decay of the
off-diagonal elements $\rho_{01}=\rho_{10}^{*}$. After preparation
of the input state, a small noisy magnetic field $\Delta B(t)$ is
superimposed onto to the static field $B_0$ that defines the
quantization axis. For this purpose an additional magnetic field
coil is placed near the trapped ion and is fed by a signal generator
producing (nearly) white noise with a Gaussian amplitude
distribution. The bandwidth of this additional noise field is
limited by a first-order filter with cut-off frequency $\omega_c=
750$ Hz. Its amplitude is experimentally controlled using a variable
attenuator.

The resonance frequency of the qubit when exposed to a magnetic field is
derived using the Breit-Rabi formula \cite{Breit31,Corney78} as
 \be
\nonumber \omega_0 (\chi) &=& \frac{1}{\hbar} E_{hfs} \sqrt{1+\chi^2} \\
   & \approx & \frac{1}{\hbar} E_{hfs} \left(1 + \frac{\chi^2}{2}\right)
 \ee
where the hyperfine splitting in zero magnetic field is denoted by
$E_{hfs}$, the scaled magnetic field is given by
\begin{equation}
 \chi\equiv \frac{\left( g_J + g_I \frac{m_e}{m_p}\right)\mu_B
 B}{E_{hfs}}\ ,
\end{equation}
the total applied magnetic field is denoted by $B$, $m_e$ and $m_p$
indicate the electron and proton masses, respectively, $g_J$ and $g_I$
are the electronic and nuclear g-factor, and $\mu_B$ is the Bohr
magneton.

The scaled magnetic field consists of two parts, $\chi_0 \propto
B_0$ and $\Delta\chi \propto \Delta B$ such that to the lowest order in
$\Delta\chi$
 \be
\nonumber \omega_0 (\chi) &\approx & \omega_0 + \frac{\omega_0}{2}
 +\omega_0\chi_0\Delta\chi \\
    &\equiv & \omega(\chi_0) + \Delta\omega
 \ee
and $\Delta\omega(t) \propto \Delta\chi(t)$. During the free precession
of the qubit its dynamics is governed by the Hamiltonian
\begin{equation}
H = \frac{\hbar}{2}\left(\omega(\chi_0) + \Delta\omega(t)\right)
\sigma_z
\end{equation}
and after transforming into a rotating frame using
$U=\exp(-\frac{i}{2}\omega(\chi_0)t\sigma_z)$ the qubit's state
evolves according to
\begin{equation}
|\psi(t)\rangle = \exp\left(-\frac{i}{2}\varphi(t)\sigma_z\right)
|\psi(t=0)\rangle \
\end{equation}
with \cite{Brouard03}
\begin{equation}
\varphi(t) = \int_0^t \Delta\omega(t') dt' \ .
\end{equation}
The magnetic field fluctuations $\Delta B(t)$ obey a Gaussian
distribution. Thus,  $\Delta\omega(t)$, $d\varphi = \Delta\omega(t)
dt$, and also $\varphi(t)$ are distributed according to a Gaussian
function. When measuring a single instance realization of the qubit
after time $t$, the off-diagonal element reads
$\rho_{01}e^{-i\varphi(t)}$ and repeating such a measurement many
times amounts to averaging over many realizations of $\varphi(t)$.

The above considerations are valid as long as the correlation time
$1/\tau \approx \unit[750]{Hz}$ of the noise field is much shorter
than the time during which the noise field is applied which is
always the case in the experiments presented here. Analytical and
numerical calculations on the dephasing of a qubit exposed to
Gaussian low-frequency noise are reported by Rabenstein {\it et al.}
\cite{Rabenstein04}. Their results are applicable to our experiment
when setting their qubit tunnel amplitude, $\Delta$  to zero
($\Delta$ corresponds to the Rabi frequency $\Omega$ in our
experiment which is zero while the noisy magnetic field is applied).
Rabenstein {\it et al.} obtain in the long-time limit ($\tau \ll
t$) for the damping factor of the off-diagonal elements
$\exp(-S_v(0) t/2)$. Here $S_v(0)$ is the spectral density at zero
frequency of the applied Gaussian noise.

As the bandwidth of the noise magnetic field is much smaller than
the Zeeman splitting of the $F=1$ state, the ionic state follows
this additional field adiabatically and transitions between
different Zeeman states are not induced by this noise field.

%%%%%%%%%%%%%%%%%%%%%%%%%%%%%%%%%%%%%%%%%%%%%%%%%%%%%%%%%%%%%%%%%%%%
%%%%%%%%%%%%%%%%%%%%%%%%%%%%%%%%%%%%%%%%%%%%%%%%%%%%%%%%%%%%%%%%%%%%
%%%%%%%%%%%%%%%%%%%%%%%%%%%%%%%%%%%%%%%%%%%%%%%%%%%%%%%%%%%%%%%%%%%%
\subsubsection{Data representation}
A quantum channel $\cE$, \ie\ the propagation of a qubit between
the initial preparation and the final measurement, is in general characterized by 12 real
parameters forming a matrix $M$ and vector $\vec{v}$. In the case of
{\it ideal} preparations and
measurements, i.e. $\varrho_{\pm k}=\ket{\pm k}\bra{\pm k}$ and
$F_{\pm j}=\frac{1}{2}(I\pm\sigma_j)=\ket{\pm j}\bra{\pm j}$, the
frequencies
 \be \nonumber f_{\pm j,\pm k}&=&\tr{F_{\pm
j}\cE[\varrho_{\pm k}]}=
\bra{\pm j}\cE[\ket{\pm k}\bra{\pm k}]\ket{\pm j}\\
\nonumber &=&
\frac{1}{4}(2\pm\tr{\sigma_j \cE[I]}\pm\pm\tr{\sigma_j\cE[\sigma_k]})\\
\nonumber &=&\frac{1}{2}[1\pm v_j\pm\pm M_{jk}]\, ,
 \ee
 and, therefore
 \be
 \label{eq:M_jk}
M_{jk}&=&2 f_{j,k} - f_{j,z} - f_{j,-z}\, , \\
v_j&=& f_{j,z}+f_{j,-z}-1\, .
 \ee
Note that inserting ``bare'' frequencies $f_{j,k}$ directly into Eq.~
\eqref{eq:M_jk} to calculate the matrix elements $M_{jk}$ and vector
elements $v_j$ is {\em not} justified when experimental
imperfections occur.

Nevertheless we shall use the matrix
  \be
 D_{jk}&=& 2 f_{j,k} - f_{j,z} - f_{j,-z}\
 \label{eq:D}
 \ee
 and the vector
 \be
 d_j&=&f_{j,z}+f_{j,-z}-1
 \ee
to represent the experimental data. Only if the test states
$\varrho_{\pm k}$ are pure and the measurements $\Sigma_j$ are sharp
(i.e., $\eta=1$), then $M_{jk}=D_{jk}$ and $v_j=d_j$. Otherwise, the
correct interpretation of frequencies $f_{\pm j,\pm k}$ lead to
different values of channel parameters $M_{jk}\neq D_{jk}$, $v_j\neq
d_j$. The precise and correct reconstruction relations shall be
presented and discussed in Section \ref{sec:III.B}.

In conclusion, for an experimentally realized quantum channel, the
experimental data {\em before} applying the appropriate channel
reconstruction procedures are represented in the form of a matrix
 \be {\cal D}=\left(\begin{array}{cccc}
d_x \quad & D_{xx} & D_{xy} & D_{xz} \\
d_y \quad & D_{yx} & D_{yy} & D_{yz} \\
d_z \quad & D_{zx} & D_{zy} & D_{zz} \\
\end{array}
\right)\, .
 \ee

%%%%%%%%%%%%%%%%%%%%%%%%%%%%%%%%%%%%%%%%%% complete tomography
%%%%%%%%%%%%%%%%%%%%%%%%%%%%%%%%%%%%%%%%%%%%%%%%%%%%%%%%%%%%%%%%%%%%
\section{Complete channel tomography}
\label{sec:III}
%%%%%%%%%%%%%%%%%%%%%%%%%%%%%%%%%%%%%%%%%%%%%%%%%%%%%%%%%%%%%%%%%%%%
In this section we shall describe several ways how to process the
experimental data. The experimental
data are always presented as frequencies of occurrences of particular
experimental outcomes. Ideally, the frequencies are equal
to probabilities predicted by the theory. Strictly speaking, this
is true if the number of runs of the experiment is infinite. Otherwise,
we must deal with statistical errors that can be eliminated by
employing specific statistical procedures. The goal of any channel-estimation procedure is to reliably
identify (given a specific figure of merit) the channel that is compatible with the measured data.

\subsection{Description of methods}
\subsubsection{Inverse linear method.}
Let us assume that  statistical errors are vanishingly small, i.e.
the frequencies are  exactly equal to the probabilities that are going
to be compared with predictions of the quantum theory. In the quantum theory
the relation between the channels and probabilities is linear and
invertible. However, the exact form depends on particular parameters
of the experiment, i.e. on a collection of test states and measurements.
We shall discuss the details later in Section \ref{sec:III.B}.

By definition of the procedure the estimated channel $\cE_{\rm est}$ is
a linear map and its Bloch sphere parametrization guarantees it is also
trace-preserving. One may think that experimental results must always
give a mapping $\cE_{\rm est}$ that is completely positive. However,
this is not necessarily the case. As we shall see the experimental data
processed in this way can indeed violate the complete positivity
constraint. If the experimental conditions are not affected by some
systematic errors, then this could happen only if the measured statistics
is not sufficiently large.

A naive method how to regularize the statistics is to add white
noise into the measured data. Let us denote by $\cA_0$ the contracting channel
maps of all input states into the maximally mixed state $\frac{1}{2}I$,
i.e. $\vec{v}_0=\vec{0}$ and $M_0=O$. Addition of the white noise channel 
corresponds to a substitution $\cE_{\rm est}\to \cE_c=c\cE_{\rm
est}+(1-c)\cA_0$. For many values of $c$ the mapping $\cE_c$ is
completely positive although $\cE_{\rm est}$ is not. It is natural
to fix the largest possible $c$ as the value of this correction
parameter and consider $\cE_c$ to be the ``regularized'' estimation of the
map. This type of regularization catches some important features of
the original map, but, in general, this method is not justified by
any statistical reasoning. It can be used as a fast test of the
quality of the experimental data.

\subsubsection{Maximum Likelihood method.}
In this approach we do not interpret frequencies as probabilities, but
instead we directly process the measured frequencies to estimate the channel.
The maximum likelihood method is a general estimation scheme
\cite{fisher,paris} that has already been considered for reconstruction of
quantum channels. It has been studied by Hradil and
Fiur\'{a}\v{s}ek \cite{fiurasek}, and by Sacchi \cite{sacchi}.

It is natural to understand an experiment as a collection of
settings $(\varrho_j,A_j)$ representing the choice of the test state
$\varrho_j$ and of the measurement $A_j$. Each measurement apparatus
is described by a positive operator valued measure (POVM) determining the
positive operators $F_{jk}$ such that $\sum_k F_{jk}=I$. The index
$k$ runs over all possible outcomes of the measurement $A_j$. The
observed frequencies $f_{jk}$ are calculated as the the fraction of
the number of ``clicks'' (events) associated with $F_{jk}$ and of the number of
experiments in which the setting $(\varrho_j,A_j)$ was used. Let us
note that $\sum_k f_{jk}=1$. For a given setting $(\varrho_j,A_j)$,
the quantum theory predicts the probabilities
 \be
p_{jk}=\tr{F_{jk}\cE[\varrho_j]}.
 \ee
The likelihood functional is defined by the formula
 \be L({\cal
E})=-\sum_{j,k} f_{jk}\log p_{jk}\, .
 \ee
The aim is to identify a physical map ${\cal E}_{est}$
that maximizes this function, i.e.
 \be
 \cE_{\rm est}=\arg\max_{{\cal E}}L({\cal E})\ .
 \ee
If the maximum is searched only among the (physically relevant) quantum channels, then the
complete positivity is guaranteed. In such case, the maximum
likelihood method cannot result in an unphysical mapping $\cE_{\rm
est}$. Since the complete positivity constraints are complicated
even for single-qubit channels the maximum likelihood optimization
problem is, in general,  complicated. Therefore complex numerical
methods must be employed.

As this variational task is usually performed numerically, a
question may arise, how to decide, if one has found a global or a
local minima. Unfortunately, there is never a certain answer for
this question. For proper data a good hint is the actual maximal
value of $L$ found during the optimization. For a given amount of
experimental data a value different in orders of magnitude from
other, similar results is suspicious. Also, from the construction of
the experiment one has an expectation for the actual result (e.g.
phase-damping channel).

For data, which (by the standard reconstruction method described in
the previous Section) lead to a proper physical reconstruction, the
maximum likelihood gives identical results. Also, as the resulting
operation in general is not on the border of the set of the CP
channels, the CP condition can be obeyed during the maximization
procedure, which speeds up the algorithm dramatically.

For data leading to unphysical results (using the inverse method)
the situation is more complicated. The best physical result is normally
on the border of the set
of CP channels, so one has to take the CP condition into account during the
whole procedure. Moreover, the resulting value of $L$ might vary significantly
depending on how much would the complete operation be non-CP. So the only
way for checking the result is the comparison with the expected type of a 
quantum channel.

\subsection{Data processing and interpretation of results}
\label{sec:III.B} In order to properly interpret the measured
experimental data we need to take into account the experimental
imprecisions leading to imperfection of the preparation process and also
the imperfections in the implementation of sharp measurements
$\sigma_x,\sigma_y,\sigma_z$. In fact, the latter imperfections must
be taken into account also in the specification of the preparation
process.

Let us express the states
\be
\varrho_{\pm k}=\frac{1}{2}(\idty +\vec{r}_{\pm k}\cdot\vec{\sigma})\, ,
\ee
corresponding to an imperfect preparation of eigenstates of
operators $\sigma_x,\sigma_y,\sigma_z$. In our particular experiment
we use only four of these states $\varrho_x,\varrho_y,\varrho_{\pm z}$.
As it was described in Section \ref{sec:Measurement}
the systems are measured by one of three observables $\Sigma_x,\Sigma_y,\Sigma_z$
described by positive operators
\be
F_{\pm j}=\eta P_{\pm j}+(1-\eta)P_{\mp j}=\frac{1}{2}[\idty\pm (2\eta-1)\sigma_j]\, ,
\ee
where $P_{\pm j}=\frac{1}{2}(\idty \pm\sigma_j)$ are eigenstates of
$\sigma_j$ corresponding to eigenvalues $\pm 1$.

The experiment consists of different settings
$(\varrho_{\pm k},M_j)$ leading to frequencies $f_{\pm j,\pm k}$
that are compared with theoretical probabilities
\be
\nonumber
p_{\pm j,\pm k}&=&\tr{F_{\pm j}\cE[\varrho_{\pm k}]} \\
\nonumber &=& \frac{1}{4}\tr{(\idty \pm(2\eta-1)\sigma_j)
\cE[\idty+\vec{r}_{\pm k}\cdot\vec{\sigma}]}\\
\nonumber &=&\frac{1}{2}
\pm\frac{2\eta-1}{4}
\left(\tr{\sigma_j\cE[\idty]}+\sum_l R_{l,\pm k}\tr{\sigma_j\cE[\sigma_l]}\right)\\
\nonumber
&=& \frac{1}{2}[1\pm (2\eta-1)(v_j+\sum_l R_{l,\pm k}M_{jl})]\, ,
\ee
where we used the definitions $R_{l,\pm k}=\tr{\sigma_l\varrho_{\pm k}}$,
$v_j=\frac{1}{2}\tr{\sigma_j\cE[\idty]}$ and
$M_{jl}=\frac{1}{2}\tr{\sigma_j\cE[\sigma_l]}$. The goal of the process
reconstruction is to determine 12 parameters $v_j$ and $M_{jk}$ describing the
quantum channel in the Bloch representation.

As we already mentioned in the experiment we use
only four test states $\varrho_x,\varrho_y,\varrho_{\pm z}$. In such case
\be
2f_{+j,+k}-f_{+j,+z}-f_{+j,-z}=(2\eta-1)\sum_{l} M_{jl} Q_{lk}\, ,
\ee
where $Q_{lk}=R_{l,+k}-\frac{1}{2}(R_{l,+z}+R_{l,-z})$. Similarly,
\be
f_{+j,+z}+f_{+j,-z}-1=(2\eta-1)[v_j+\sum_l M_{jl}q_l]\, ,
\ee
where $q_l=\frac{1}{2}(R_{l,+z}+R_{l,-z})$. The matrix $Q$ and vector $\vec{q}$
are fixed by performing the state reconstruction experiments
of the input test states $\varrho_{\pm k}$. Since the experimental data
are presented in the form of the matrix $D_{jk}=2f_{+j,+k}-f_{+j,+z}-f_{+j,-z}$
and the vector $d_j=f_{+j,+z}+f_{+j,-z}-1$ it follows that
\be
\label{eq:M}
M_{jl}&=&\frac{1}{2\eta-1}\sum_k D_{jk} Q_{kl}^{-1}\, ;\\
\label{eq:v}
v_j&=&\frac{1}{2\eta-1}d_j-\sum_{l} M_{jl} q_l\, .
\ee
Let remind us that this choice of the data representation is motivated by
the property that if the test states are pure
[i.e., $\varrho_{\pm k}=\frac{1}{2}(\idty\pm\sigma_k)$] and the measurements are sharp
($\eta=1$), then $v_j=d_j$ and $M_{jl}=D_{jl}$.

The preparation of the input test states is described by the parameters
\be
R_{j,\pm k}=\tr{\sigma_j\varrho_{\pm k}}=
\frac{1}{2\eta-1}(f^0_{+j,\pm k}-f^0_{-j,\pm k})\, ,
\ee
where $f^0_{\pm j,\pm k}$ are the observed frequencies for the experiment
estimating the test states $\varrho_{\pm k}$. Inserting this
formula into the expressions for $Q$ and $\vec{q}$ we find
\be
\nonumber
Q_{jk}&=&\frac{1}{2\eta-1}(2f^{0}_{j,+k}-f^0_{j,+z}-f_{j,-z}^0)=\frac{1}{2\eta-1}
D_{0,jk}\, ,\\
\nonumber
q_j&=&\frac{1}{2\eta-1}(f^0_{j,+z}+f^0_{j,-z}-1)=\frac{1}{2\eta-1}d_{0,j}\, .
\ee
That is, $Q$ and $\vec{q}$ are proportional to data matrices
determined by the preparation procedures of the test states. Hence
the data collected in the estimation of the test states can be
conveniently expressed in the same form as the data collected in the case of
channel estimation procedure. However, the data set matrix
${\cal D}_0=(\vec{d}_0,D_0)$ does not have to possess the channel
requirements on ${\cal D}$, i.e. it does not have
to be completely positive.

In the considered experiment
\be
\nonumber
{\cal D}_0=\left(\begin{array}{cccc}
 0.031 &  0.805 & 0.004 &  0.016 \\
 0.011 & -0.016 & 0.784 & -0.055 \\
-0.004 & -0.015 & 0.018 &  0.803
\end{array}\right)\; ,
\ee
and the Bloch vectors of actual test states read
\be
\nonumber \vec{r}_{+x}&=&\frac{1}{2\eta-1}(0.836,-0.005,-0.019)\, ,\\
\nonumber \vec{r}_{+y}&=&\frac{1}{2\eta-1}(0.035,0.795,0.014)\, ,\\
\nonumber \vec{r}_{+z}&=&\frac{1}{2\eta-1}(0.047,-0.044,0.799)\, ,\\
\nonumber \vec{r}_{-z}&=&\frac{1}{2\eta-1}(0.015,0.066,-0.807)\, .
\ee
The above vectors are the columns of the matrix $R_{l,\pm k}$. The parameter
$\eta$ is determined experimentally and reads $\eta=0.988$. The fidelities
$\bra{\pm k}\varrho_{\pm k}\ket{\pm k}$ of preparation
of pure input test states $\ket{x}\bra{x}$, $\ket{y}\bra{y}$,
$\ket{z}\bra{z}$, $\ket{-z}\bra{-z}$ are
0.928, 0.907, 0.909, 0.913, respectively. These particular values
shall be used in the process tomography procedure using the formulas
given by Eqs.\eqref{eq:M} and \eqref{eq:v}.

In summary, given a data set
${\cal D}=(\vec{d}, D)$, the channel $\cE$ can be very
conveniently determined by using the following matrix relations
replacing the formulas \eqref{eq:M} and \eqref{eq:v}
 \be
 \label{eq:E(D)}
 \cE=\Phi_\eta{\cal D}{\cal D}_0^{-1}{\Phi}^{-1}_\eta=\left(
\begin{array}{cc}
1 & \vec{0} \\
\vec{v} & M
\end{array}
\right)\, ,
 \ee
where
\be
\nonumber
\Phi_\eta=\left(
\begin{array}{cc}
2\eta -1 & \vec{0} \\
\vec{0} & I
\end{array}
\right)\,, \quad
{\cal D}^{-1}_0=\left(
\begin{array}{cc}
1 & \vec{0} \\
-D_0^{-1}\vec{d}_0 & D^{-1}_0
\end{array}
\right)\, .
 \ee
%%%%%%%%%%%%%%%%%%%%%%%%%%%%%%%%%%%%%%%%%%%%%%%%%%%%%%%%%%%%
%%%%%%%%%%%%%%%%%%%%%%%%%%%%%%%%%%%%%%%%%%%%%%%%%%%%%%%%%%%%
\section{Phase damping quantum channels}\label{sec:IV}
The action of the phase damping quantum channel on the Bloch vector 
is described by the matrix
\begin{equation}\label{eq:phasedamping}
\cE_\lambda:\vec{r}\to\vec{r}^\prime = \left(\begin{array}{ccc}
\lambda&0&0\\
0&\lambda&0\\
0&0&1
\end{array}\right)
\vec{r}
\end{equation}
with the damping parameter $|\lambda| \le 1$. It reduces the
$x$th and $y$th components of the Bloch vector while the $z$th component
remains unaffected.

As described in Section \ref{sec:phase_damping} a phase damping
quantum channel is realized by applying a normally distributed noise
magnetic field between the preparation and the measurement stages. This noise field
is applied for a duration that equals a multiple of the precession
time $T_P$ with an amplitude determined by a variable attenuator.
The noise magnetic field changes the precession frequency of the ion
by a small random amount and therefore adds noise to the phase of
the qubit. Upon averaging over many realizations this reduces the
$x$- and $y$-components of the Bloch vector.

The damping parameter $\lambda$ describes the amount of phase
damping that occurs:
\begin{equation}\label{eq:lambda_exp}
\lambda = \exp\left[-S_v(0)\frac{t}{2}\right]
\end{equation}
with $t$ being the duration
for which the noise magnetic field is applied and $S_v(0)$ is the
spectral density of the applied Gaussian noise at zero frequency.

\begin{figure}
\includegraphics[width=\hsize]{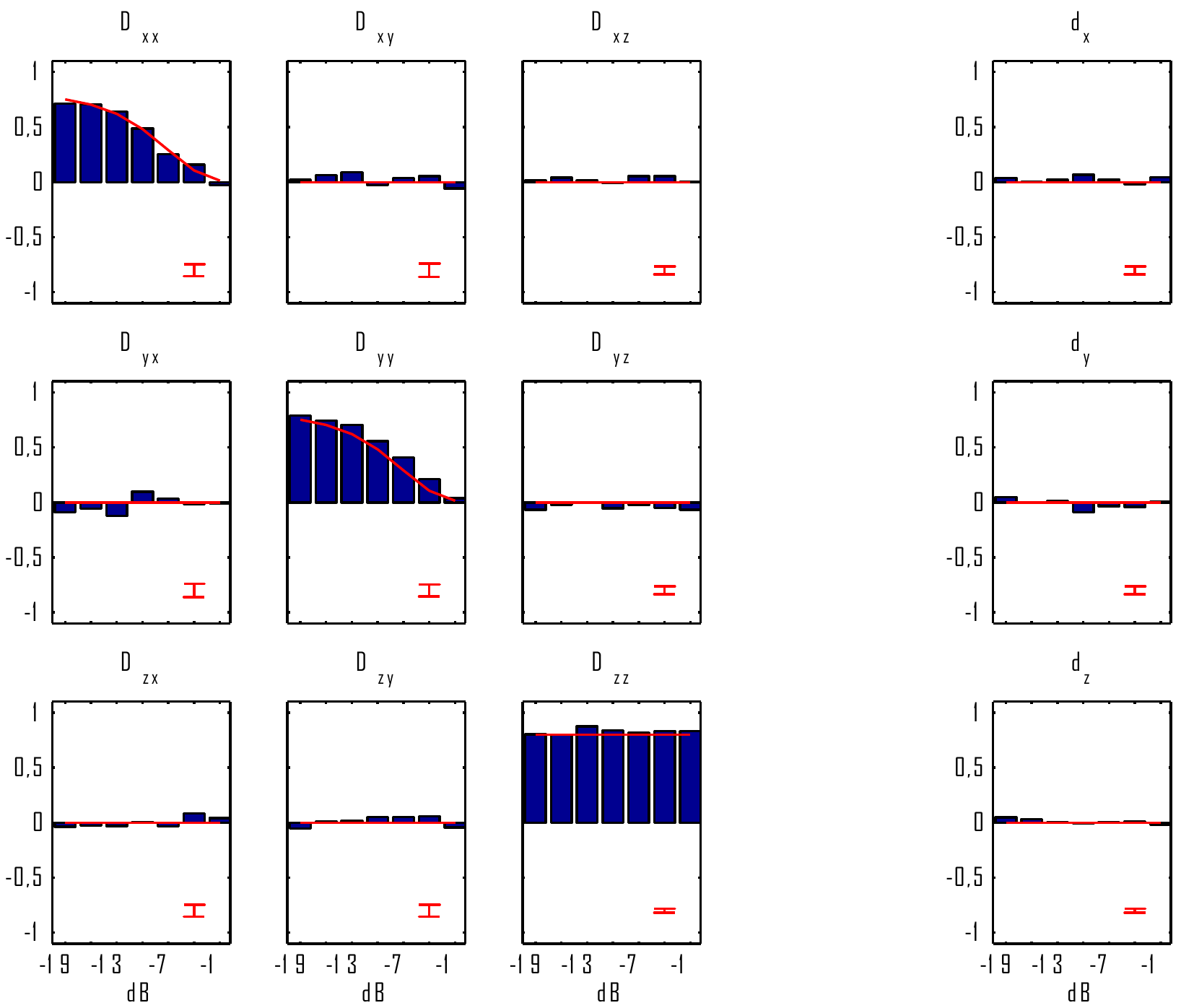}
\caption{ \label{fig:phasedamping_amplitude} (color online). Phase-damping quantum
channels. Shown are the experimentally determined values of the
matrix $M=D$ and the vector $\vec{v}=\vec{d}$ as a bar plot. The
amount of phase damping is varied by changing the amplitude
(indicated on the x-axes) of the noise magnetic field that is
applied for a fixed time of $2T_P= 4\pi / \delta = \unit[21.6]{ms}$.
The diagonal elements $M_{xx}$ and $M_{yy}$ are fitted as a function
of the noise amplitude employing Eq. \protect\eqref{eq:lambda_exp} (solid
lines). The error bar in the lower right corner of the diagram shows
the averaged error of the experimental values in this plot.  }
\end{figure}

In order to vary the damping parameter $\lambda$ experimentally, the
amplitude of the noise magnetic field is changed while the duration
is fixed at $2T_P= 4\pi / \delta = \unit[21.6]{ms}$ for the data
shown in Fig. \ref{fig:phasedamping_amplitude}. The relative noise
amplitude is indicated on the x-axis for each matrix and vector
element displayed in Fig. \ref{fig:phasedamping_amplitude}. Already
in this figure we can see the expected pattern of a phase damping
channel. That is, the off-diagonal elements $D_{jk}$ and values of
$d_j$ almost vanish, while the element $D_{zz}$ remains almost
constant and the values of $D_{xx},D_{yy}$ are exponentially
decreasing to zero as the amplitude of the noise magnetic field is
increasing: The solid (red) line in Fig. \ref{fig:phasedamping_amplitude}
indicates a fit with an exponential decay using Eq.
\eqref{eq:lambda_exp} where $S_v^0(0)$ at a relative amplitude of 0
dB is used as a free parameter and $S_v(0)$ varies according to
$S_v(0)= S_v^0(0)\cdot 10^{s/10}$.  Here, $s$ indicates the relative
noise amplitude indicated in dB in Fig.
\ref{fig:phasedamping_amplitude} and the time is fixed at $t=2T_p$.

The fact that $D_{zz}<1$ is consistent with the effective initial
preparation of the qubit in a mixed state and its imperfect
detection. This matrix element is expected to remain unaffected by
the applied noise, since the noise doesn't induce transitions
between the qubit states. This indeed is found to be the case.

The error bars shown in Fig. \ref{fig:phasedamping_amplitude}
indicate a statistical error originating from a finite number of
measurements that go into the determination of the relative
frequencies and consequently into the matrix elements calculated
according to Eq.\eqref{eq:D}.

\subsection{Estimation without prior knowledge}
 \label{sec:E_j}
For example, for the data set corresponding to $s=-10$~dB
represented by the fourth bar in each individual graph of Fig.
\ref{fig:phasedamping_amplitude},
\begin{equation*}
{\cal D}_4=\left(
\begin{array}{cccc}
1 & 0 & 0 & 0 \\
-0.09 &  0.56 & -0.03 & 0.05 \\
 0.07 & -0.10 & 0.49 & 0.00 \\
-0.01 & -0.05 & -0.01 & 0.84
\end{array}
\right)\, ,
\end{equation*}
and the linear inverse reconstruction method using Eq.~
\eqref{eq:E(D)} results in the mapping
\begin{equation*}
\cE_{4}=\left(
\begin{array}{cccc}
1 & 0 & 0 & 0 \\
-0.12 &  0.69 &  -0.04 & 0.05 \\
0.06 &  0.13   & 0.62  & 0.04 \\
0.00 & -0.05   & -0.04  & 1.04
\end{array}
\right)\, .
\end{equation*}
Since $M_{zz}>1$ it follows that the whole mapping is not positive,
hence the reconstruction gives an unphysical result. We find that
this feature of ``unphysicality'' is typical for all values of $s$.
However, this is not entirely unexpected, because phase damping
channels are on the boundary between positive and non-positive maps.
That is, for each phase damping channel $\cE_\lambda$ there exists a
non-positive linear map which is arbitrarily close to $\cE_\lambda$.
In fact, in our case the violation of the complete positivity is
within the statistical errors. The used statistics is relatively
small. Each experiment was repeated 100 times.

\begin{figure}
\includegraphics[width=\hsize]{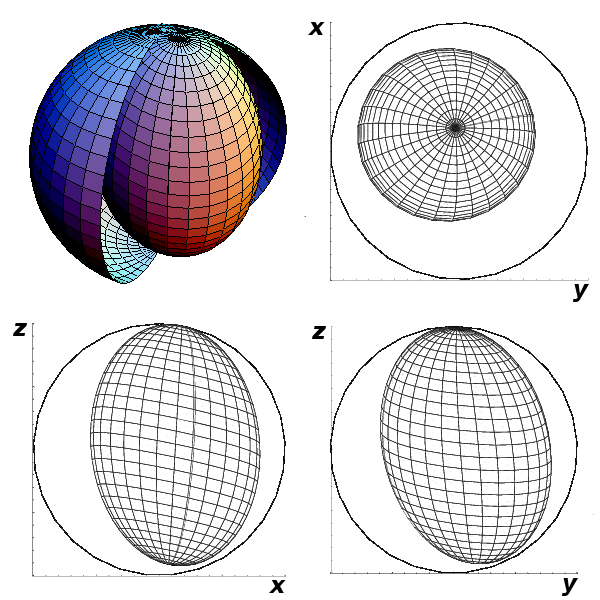}
\caption{(color online). The visualization of the action of the estimated channel $\cE^{\rm est}_4$ on
the Bloch sphere. The original Bloch sphere corresponding to input states is transformed
by channel $\cE^{\rm est}_4$ into  the ``ellipsoid'' corresponding to the state space of output states.
We also present projections of both the original Bloch sphere and the output ellipsoid onto the planes $xy,xz,yz$. }
\label{fig:16_ff}
\end{figure}

Employing the maximum
likelihood principle for the data taken at $s=-10$~dB that
was considered already above we now get
\begin{equation*}
\cE_{4}^{\rm est}=\left(
\begin{array}{cccc}
1 & 0 & 0 & 0 \\
-0.17 &  0.60 &  -0.13 & 0.16 \\
0.04 &  0.15   & 0.60  & 0.00 \\
0.00 & -0.15   & 0.03  & 0.91
\end{array}
\right)\, .
\end{equation*}
We see that this (estimated physical) mapping is not exactly the phase damping channel
(see Fig.\ref{fig:16_ff}), but
the obtained precision is in accordance with the size of the statistical
sample.

%%%%%%%%%%%%%%%%%%%%%%%%%%%%%%%%%%%%%%%%%%%%%%%%%%%%%%%%%%%%%%%%
%%%%%%%%%%%%%%%%%%%%%%%%%%%%%%%%%%%%%%%%%%%%%%%%%%%%%%%%%%%%%%%%%%%%
\subsection{Estimation of the phase damping parameter}
%%%%%%%%%%%%%%%%%%%%%%%%%%%%%%%%%%%%%%%%%%%%%%%%%%%%%%%%%%%%%%%%%%%%
So far we reconstructed the quantum channel without using any
information on the particular physical model of the the experiment.
In this part we shall assume that nontrivial {\it a priori} information on
the estimated channel is available. In particular, we shall consider
the channel described and reconstructed above and assume prior knowledge that the
channel describes a pure phase damping channel $\cE_\lambda$. Let us
note, that the results of Section \ref{sec:IV} do no entirely
justify such an assumption, because the implemented channels are not
precisely the phase damping channels. Nevertheless, our goal is to
present different methods how to determine the phase damping
parameter $\lambda$, which is probably the most interesting
parameter of any decoherence evolution, because it illustrates how
fast the ``quantumness'' of a given quantum system is deteriorated.

Firstly, we shall employ the
maximum likelihood method constrained to phase damping channels
only. In this case the optimization of the likelihood is constrained
only to channels $\cE_\lambda$, that is,
\be
\label{eq:lambda_est}
\lambda_{\rm
est}=\arg\max_\lambda \sum_{jk} f_{\pm j,\pm k}\log p_{\pm j,\pm k}
(\lambda)\, , \ee where $p_{jk}(\lambda)=\tr{F_{\pm
j}\cE_\lambda[\varrho_{\pm k}]}$.

In the second approach we shall also assume the same form of the
channel. Now the parameter $\lambda$ reads
\be \label{eq:lambda}
\lambda=M_{xx}=\tr{\sigma_x\cE_\lambda[\sigma_x]}=
\frac{\tr{\sigma_x \cE_\lambda[\varrho]}} {\tr{\sigma_x\varrho}}
 \ee
for an arbitrary test state $\varrho$. Alternatively, one
can replace $\sigma_x$ by any other operator
$\vec{t}\cdot\vec{\sigma}$ with a vanishing $z$-component of
$\vec{t}$, i.e. orthogonal to $\sigma_z$ determining the decoherence
basis. Since the relative statistical error is largest for
off-diagonal elements, in order to fix some value of $\lambda$ we
shall use the average value of $M_{xx}$ and $M_{yy}$, i.e.
$\overline{\lambda}=(M_{xx}+M_{yy})/2$.

\begin{figure}
\includegraphics[width=8cm]{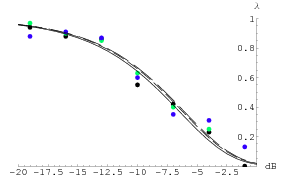}
\caption{(color online). Dependence of the decoherence rate $\lambda$ for values presented in
Tab.~\ref{table:lambda}. The interpolation gives us the values of
$S_v^0(0) = 0.41 \unit{ms}^{-1}$,  $S_v^0(0) = 0.38 \unit{ms}^{-1}$,
$S_v^0(0) = 0.37 \unit{ms}^{-1}$ for $\overline{\lambda}$, $\overline{\lambda}_{ML}$,
$\lambda_{\rm est}$, respectively.
}
\label{fig:lambda}
\end{figure}

In Tab.~\ref{table:lambda} and Fig.~\ref{fig:lambda} we present the
estimates of the value of $\lambda$ using three methods: i) the
average $\overline{\lambda}=(M_{xx}+M_{yy})/2$ for $\cE_j$ using the
linear inverse reconstruction (compare Section \ref{sec:E_j}), ii)
the average $\overline{\lambda}_{ML}$ for $\cE_{j}^{est}$ obtained
from maximum likelihood, and iii) $\lambda_{\rm est}$ obtained from the
constrained maximum likelihood [see Eq.\eqref{eq:lambda_est}].

\begin{table}[h]
\begin{tabular}{|c|c|c|c|c|c|c|c|}
\hline
setting & -19dB & -16dB & -13dB & -10dB & -7dB & -4dB & -1dB
\\ \hline
$\overline{\lambda}$ &  0.94 & 0.88 & 0.85 & 0.66 & 0.42 & 0.23 & 0.00
\\ \hline
$\overline{\lambda}_{ML}$ & 0.88 & 0.91 & 0.87 & 0.60 & 0.35 & 0.31 & 0.13
\\ \hline
$\lambda_{\rm est}$ &  0.97 & 0.90 & 0.85 & 0.63 & 0.40 & 0.25 & -0.01
\\
\hline
\end{tabular}
\caption{Estimation of phase damping rate $\lambda$ obtained with the three
different methods: i) $\overline{\lambda}=(\tr{\sigma_x{\cal
E}_{4}[\sigma_x]}+\tr{\sigma_y{\cal E}_{4}[\sigma_y]})/2$, ii)
$\overline{\lambda}_{ML}=(\tr{\sigma_x{\cal E}^{\rm
est}_{4}[\sigma_x]}+\tr{\sigma_y{\cal E}^{\rm
est}_{4}[\sigma_y]})/2$, and iii) $\lambda_{\rm est}$ is defined by
Eq.\eqref{eq:lambda_est}. } \label{table:lambda}
\end{table}

All three methods for evaluation of $\lambda$ must give the same
value in the limit of infinite statistics (providing that the
implemented channels are precisely the phase damping channels). For small
statistical samples the first method is, in general, very
inappropriate, because it can give even unphysical values. On the
other hand the values $\overline{\lambda}_{ML}$ and
$\overline{\lambda}_{\rm est}$ can give different values even for
infinite statistics. If this is the case, then we must conclude that
the phase damping channels are not implemented and the value
$\lambda$ does not have exactly the desired meaning. In our case the
statistical errors in specification of $\lambda$ are $\pm 0.1$.
Therefore, we can conclude that all three values are approximately
the same. Since the differences between $\lambda_{\rm est}$ and
$\overline{\lambda}_{ML}$ are relatively small (in the context of
statistical errors) we can conclude that phase damping channels are
realized with quite good accuracy although a precise quantitative
specification would require more experimental runs.

%%%%%%%%%%%%%%%%%%%%%%%%%%%%%%%%%%%%%%%%%%%%%%%%%%%%%%%%%%%%%%%%
\section{Phase damping quantum channels with change of basis}
The phase damping quantum channel shown in the previous paragraph
acts in the $(x,y)$ plane of the Bloch sphere. The phase damping can be
applied in a different plane, if the qubit state is rotated prior to
application of the noise magnetic field and rotated back afterwards.
Here, phase damping in a plane rotated around the $x$-axis spanned
by the Bloch vectors $(0,\sin\theta,\cos\theta)$ with $\theta=\pi/4$
and $(1,0,0)$ is examined. For arbitrary $\theta$, such rotated phase
damping is described by the Bloch vector transformation
\begin{eqnarray}\label{eq:phasedamping_wcob}
\vec{r}^\prime=
\left(\begin{array}{ccc} \lambda &0
&0\\
0& \lambda\cos^2\theta +\sin^2\theta &
(\lambda-1)\cos\theta\,\sin\theta\\
0& (\lambda-1)\cos\theta\,\sin\theta&
\lambda\sin^2\theta+\cos^2\theta\\
\end{array}\right)
\vec{r}
\end{eqnarray}
with $0 \le \lambda \le 1$ the damping parameter as in Eq.~
\eqref{eq:phasedamping}.

\begin{figure}
\includegraphics[width=\hsize]{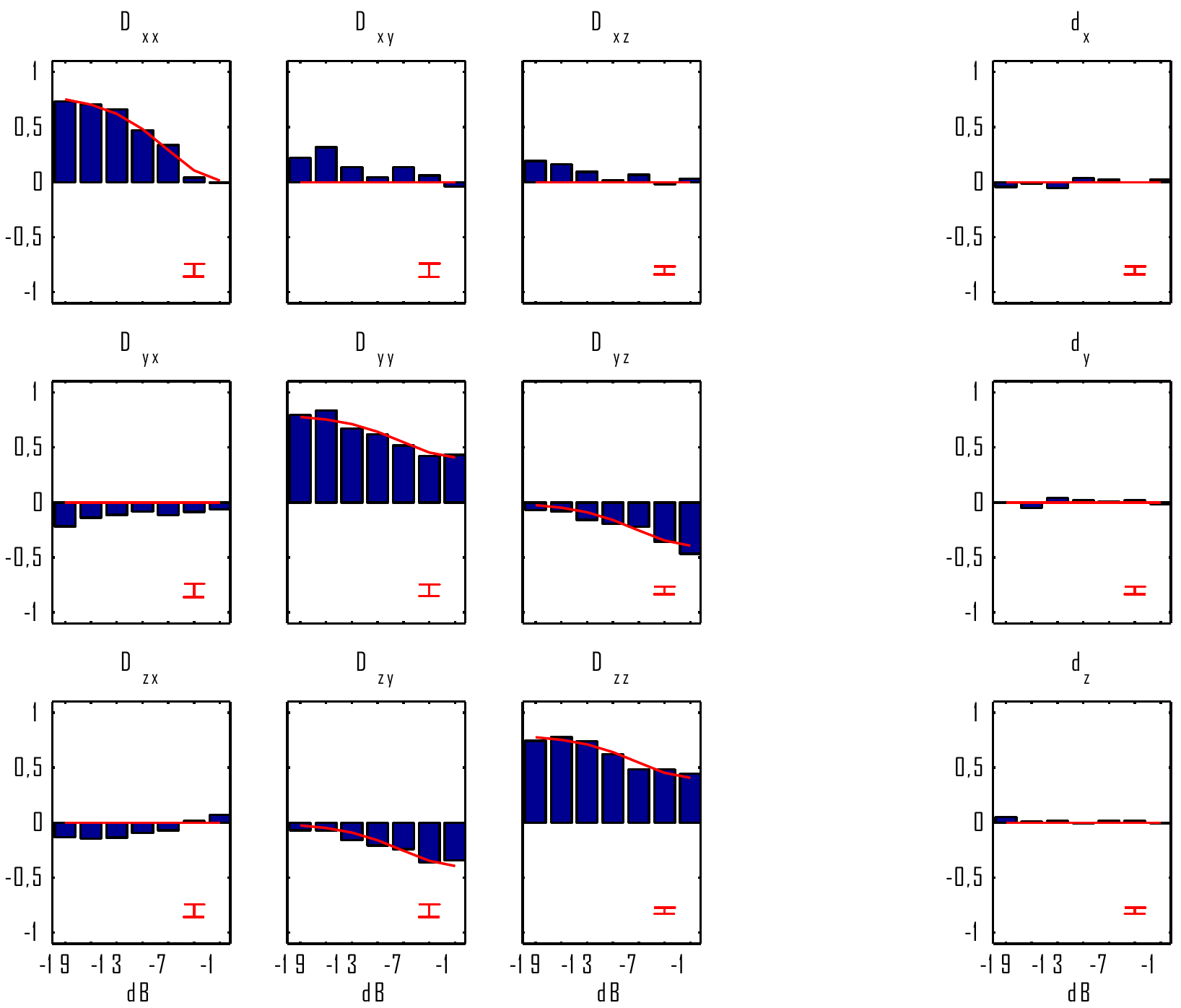}
\caption{\label{fig:phasedamping_wcob_a} (color online).  Quantum channels where
phase damping acts in the plane spanned by the Bloch vectors
$(0,\sin\pi/4,\cos\pi/4)$ and $(1,0,0)$. Shown are the
experimentally determined values of the matrix $M=D_0$ and the
vector $\vec{v}=\vec{d}_0$ in a form of a bar plot as a function of the relative
amplitude of the applied noise field. The solid lines are the result
of a fit using Eqs. \eqref{eq:phasedamping_wcob}\ and
\eqref{eq:lambda_exp}. The noise magnetic field is applied for a fixed
time of $2T_P= 4\pi / \delta = \unit[21.6]{ms}$. The error bar in
the lower right corner of the diagram shows the averaged error of
the experimental values in this plot.}
\end{figure}

Fig.~\ref{fig:phasedamping_wcob_a}\ shows the experimental results
for a phase damping channel with varying amplitude of the noise
field and fixed $\theta=\pi/4$. The solid lines indicate a fit of
the data using Eqs.~\eqref{eq:phasedamping_wcob}\ and
\eqref{eq:lambda}\ as described above.

As in the previous case, also for this type of channels the basic
features are already seen from the data presentation in
Fig.\ref{fig:phasedamping_wcob_a}. In contrast to the ``pure'' phase
damping channels shown in Fig.~\ref{fig:phasedamping_amplitude},
for this channel the off-diagonal elements do not vanish.
Furthermore, the z-component is damped. In this set of data the
values for $M_{\text{yy}}$ and $M_{\text{zz}}$ are almost equal
because of the particular choice of the rotated basis.

The channel was analyzed in a similar way as the phase damping
channel in the previous section. Let us present in detail the
reconstruction based on an experiment where the same amount of phase
damping was present as in the case of phase damping in the
$xy-$plane, i.e. $s=$\unit[-10]{dB}. For the experimental setting
\unit[-10]{dB} the data matrix reads
\begin{equation*}
{\cal D}_4=\left(
\begin{array}{cccc}
1 & 0 & 0 & 0 \\
-0.04 &  0.47 & -0.08 & -0.09 \\
 0.02 &  0.05 &  0.62 & -0.20 \\
 0.00 &  0.02 & -0.19 & 0.62
\end{array}
\right)\,.
\end{equation*}
The inverse reconstruction method gives
\begin{equation*}
{\cal E}_4=\left(
\begin{array}{cccc}
1 & 0 & 0 & 0 \\
 0.02 &  0.58 & -0.10 & -0.14 \\
 0.00 &  0.07 &  0.79 & -0.21 \\
 0.00 &  0.03 & -0.26 & 0.76
\end{array}
\right)\,,
\end{equation*}
and the maximum likelihood estimation results in channel
\begin{equation*}
{\cal E}_4^{\rm est}=\left(
\begin{array}{cccc}
1 & 0 & 0 & 0 \\
 0.01 &  0.44 & -0.00 & -0.14 \\
 0.04 &  0.12 &  0.69 & -0.31 \\
 0.04 &  0.02 & -0.24 & 0.72
\end{array}
\right)\,.
\end{equation*}

The constrained maximum likelihood method applied to channels of the
form \eqref{eq:phasedamping_wcob} results in the estimates
on the phase damping rate and the rotation axis angle presented in Tab.~II.
\begin{center}
\begin{table}[h]
\begin{tabular}{|r|c|c|c|c|c|c|c|}
\hline
setting & -19dB & -16dB & -13dB & -10dB & -7dB & -4dB & -1dB \\
\hline
$\theta_{\rm est}$ & 32$^\circ$ & 32$^\circ$ & 24$^\circ$ & 42$^\circ$ & 41$^\circ$ & 43$^\circ$ & 44$^\circ$  \\
\hline
$\lambda_{\rm est}$ & 0.88 & 0.90 & 0.74 & 0.50 & 0.38 & 0.18 & -0.04 \\
\hline
\end{tabular}
\caption{Estimation of the phase damping rate $\lambda_{\rm est}$ and the rotation axis angle $\theta_{\rm est}$
based on the constrained maximum-likelihood method.
} \label{table:table2}
\end{table}
\end{center}

In order to fix the parameters $\lambda,\theta$ experimentally it is
sufficient to specify any pair of nonzero elements of the matrix in
Eq.\eqref{eq:phasedamping_wcob}. For example, we can use
$\ket{+z}\bra{+z}$ as the test state and measure
$\sigma_y,\sigma_z$. From the estimated matrix elements the values
can be easily calculated. However, in our case we used all the
observed values as an input into the constrained maximum likelihood
estimation that determines the values of $\lambda_{\rm est}$ and
$\theta_{\rm est}$. As expected, the value of the phase damping rate
for $s=-10$dB is in accordance with the corresponding values of
$\lambda$ in Tab.~\ref{table:lambda} within the statistical
uncertainty.

The correct value (specified by the experimental setup) of the angle
$\theta$ is $45^\circ$. For weak
damping (i.e., small values of added noise and consequently large
values of $\lambda$) the estimated angle deviates from the
angle used in the experiment more than for a large damping. However, this
deviation is still within the errors of the maximum likelihood
estimates that are $\pm 10^\circ$. The larger deviation when low
damping is applied can be rationalized as follows: The fact that the
phase damping does not act in the $xy$-plane becomes apparent in the
increase of the off-diagonal elements ${\cal D}_{yz}$ and ${\cal
D}_{zy}$ of the matrix ${\cal D}_0$ only for relatively large
damping (compare Fig. \ref{fig:phasedamping_wcob_a}). For small
damping these off-diagonal elements are small with a relatively
large statistical error bar, which makes it difficult to accurately
estimate for each {\it individual} channel the angle $\theta$ by
which the plane of phase damping is rotated.

The solid line in Fig.~\ref{fig:phasedamping_wcob_a}
indicates a fit (as opposed to a channel reconstruction) of a given
matrix element ${\cal D}_{ij}$ using Eqs.
\eqref{eq:phasedamping_wcob}\ and \eqref{eq:lambda}. Here, the fit
includes $\theta$ as a free parameter and takes into account {\it
all} available results for ${\cal D}_{ij}$ (i.e., for different
strengths of damping) thereby presuming that the channels are
characterized by the same angle $\theta$ irrespective of the
strength of damping (which indeed was experimentally realized).

%%%%%%%%%%%%%%%%%%%%%%%%%%%%%%%%%%%%%%%%%%%%%%%%%%%%%%%%%
\section{Polarization rotating and phase damping quantum channels}
\begin{figure}
\includegraphics[width=\hsize]{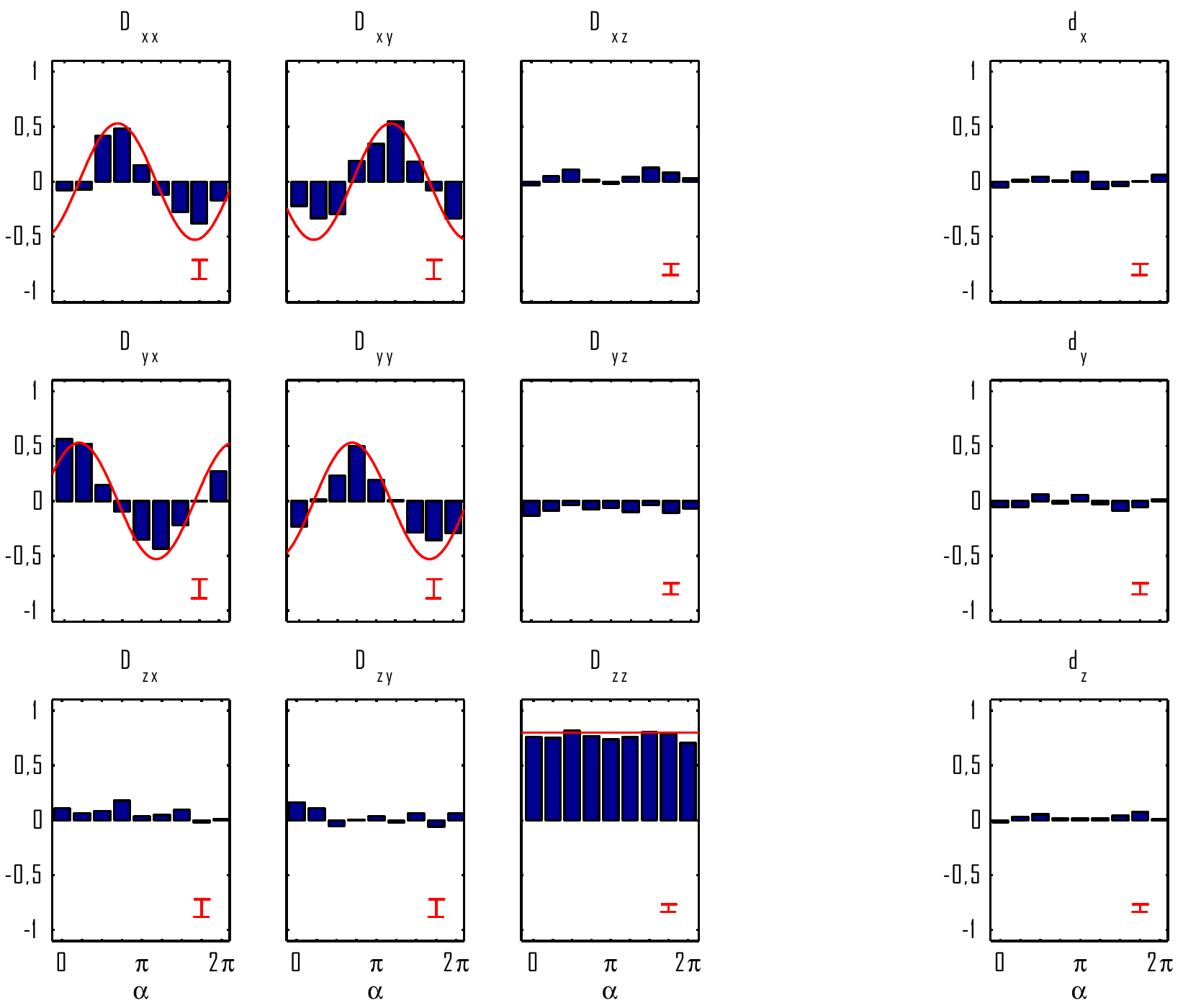}
\caption{\label{fig:combined} (color online). Quantum channels corresponding to combined phase damping
and polarization rotation. Shown are experimentally
determined values of the matrix $\Matrix$ and the vector $\Vector$
as a bar plot and the values according to the Eqs. \eqref{eq:combined}\
and \eqref{eq:lambda}\ are represented as a solid (red) line. The
relative amplitude of the noise magnetic field is \unit[-10]{dB}.
The error bar in the lower right corner of the diagram shows the
averaged error of the experimental values in this plot. }
\end{figure}

%%%%%%%%%%%%%%%%%%%%%%%%%%%%%%%%%%%%%%%%%%%%%%%%%%%%%%%%%
In a further experiment we combined the phase damping channel with a
unitary channel rotating the state space around the $z$ axis by an
angle $\alpha$. Such polarization rotating channel is realized in
our experiment by inserting a pause between the preparation of the
qubit and its measurement whose length is a fraction of the
precession time $T_P$. The combined quantum channel realized here
propagates the qubit first through a phase damping quantum channel
according to Eq.~ \eqref{eq:phasedamping} with a relative
amplitude of the noise field of  \unit[-10]{dB} and then through a
phase rotating quantum channel whose rotation angle $\alpha$ is
varied. The combined action on the Bloch vector is given by the
transformation
\begin{equation}\label{eq:combined}
\vec{r}^\prime = \left(\begin{array}{ccc}
\phantom{-}\lambda\cos\alpha & \lambda\sin\alpha & 0\\
-\lambda\sin\alpha & \lambda\cos\alpha & 0\\
0 & 0 & 1
\end{array}\right)
\vec{r}\, .
\end{equation}
In Fig.~\ref{fig:combined}\ the experimental results for that
quantum channel are shown.

One would expect that  the zero pause should correspond
to the rotation angle $\alpha=0$ for which the diagonal matrix
elements are all at their maximum. This is not the case for the data
displayed in Fig. \ref{fig:combined} where the offset of the phase
rotation angle $\alpha$ is due to an additional dc magnetic field
that was applied simultaneously with the ac noise field.

Tab.~III contains the results of the constrained
maximum likelihood estimation of the relevant parameters.
\begin{center}
\begin{table}[h]
\begin{tabular}{|r|c|c|c|c|c|c|c|c|c|}
\hline
$\alpha_{\rm est}$ & 255$^\circ$ & 271$^\circ$ & 325$^\circ$ & 22$^\circ$ & 69$^\circ$ & 90$^\circ$ & 147$^\circ$ & 171$^\circ$ & 217$^\circ$ \\
\hline
$\lambda_{\rm est}$ & 0.56 & 0.58 & 0.58 & 0.66 & 0.55 & 0.58 & 0.41 & 0.53 & 0.45\\
\hline
\end{tabular}
\caption{Estimation of the phase damping rate $\lambda_{\rm est}$ and the rotation  angle $\alpha_{\rm est}$
based on the constrained maximum-likelihood method.
} \label{table:table3}
\end{table}
\end{center}
Let us note that the values of the damping parameter are in an
approximate accordance with the results obtained in the estimations
of the previous channels, for which we have seen that the damping
parameter for \unit[-10]{dB} is around $\lambda=0.6$. In particular,
the average gives $\lambda=0.54$. Moreover, the rotation angles in
the experiment are chosen such that the differences of subsequent
angles should be $45^\circ$. According to above estimations the
average difference is $40^\circ$.

%%%%%%%%%%%%%%%%%%%%%%%%%%%%%%%%%%%%%%%%%% summary
%%%%%%%%%%%%%%%%%%%%%%%%%%%%%%%%%%%%%%%%%%%%%%%%%%%%%%%%%%%%%%%%%%%%
\section{Conclusion}
%%%%%%%%%%%%%%%%%%%%%%%%%%%%%%%%%%%%%%%%%%%%%%%%%%%%%%%%%%%%%%%%%%%%
Quantum process estimation is a necessary tool for characterization of
dynamics of physical systems. It can also be used for improvement of efficiency
of quantum information processing. We have exposed individual trapped ions to
engineered quantum channels and explored various methods for a
reconstruction of the action of these channels.
This reconstruction takes into account imperfect
experimental conditions, namely the imperfect preparation of the test
states (that results in a mixed state), a finite detection
efficiency and a bias in the detection efficiency for different
states.

We have created qubit quantum channels with variable phase
damping and fully reconstructed them using the linear inverse method
and the maximum likelihood estimate. Alternatively, if it is {\it a priori} known
that the quantum channel only leads to phase damping, then the
estimation of a single parameter is sufficient. In the latter case a
constrained maximum likelihood method leads to good results with a
minimal number of measurements.

A quantum channel with damping in an arbitrary plane
through the Bloch sphere that contains the origin may also be
realized and is exemplified here for a particular rotation angle
that determines this plane. The full reconstruction of this channel
is performed without prior knowledge using the linear inverse
and the maximum likelihood methods. Here, too, the maximum likelihood
method under constraints gives good results with a significantly 
reduced number of measurements when prior knowledge about the
channel's action is assumed and used. Furthermore, using the same methods
the phase damping accompanied by a polarization rotation is estimated.

The finiteness of experimental statistics affects the precision of our
estimates. On average the precision of matrix elements is $\pm 0.1$.
From the observed dependence of the phase-damping rate $\lambda$ on
parameter $s$ (see Fig.\ref{fig:lambda}) we can determine the
constant $S_v^0(0)$. In particular, fitting the estimated values of
$\lambda$ we get $S_v^0(0)=0.38 \unit{ms}^{-1}$.

The implementation of the phase damping channel is based on a clear
physical picture based on our knowledge from atomic physics.
However, such knowledge is not really used in the complete
tomography methods used in this paper. The adopted approach is to consider the
experiment as an unknown black box transforming states. Therefore,
comparing the estimation with the theoretical expectations gives us
nontrivial information about the validity of our assumptions and
understanding of the physical situation.

For the constrained maximum likelihood estimation we assume that the
channel belongs to a family of phase damping channels characterized
by a single parameter. For infinite statistics any difference
between unconstrained and constrained maximum likelihood estimations
would imply that the model and the experiment do not fit perfectly.
Therefore, the distance $d({\cal E}^{\rm est},{\cal E}_{\lambda_{\rm
est}})$ can provide the quantification of the agreement of the
model and the experiment.
Let us use as a figure of merit the {\it process fidelity} \cite{nielsen}
\be
F({\cal E}_1,{\cal E}_2)=\tr{\sqrt{\sqrt{\omega_1}\omega_2\sqrt{\omega_1}}}\, ,
\ee
where $\omega_j={\cal E}_j\otimes{\cal I}[\Psi_+]$ and $\Psi_+$
is a projection onto the maximally entangled state. This positive
functional equals to unity if and only if the two processes under consideration are the same
and it is less than unity (though non-negative) otherwise. For phase damping channels
we find that
\be
F({\cal E}^{\rm est},{\cal E}_{\lambda_{\rm est}})\approx 0.97\, .
\ee
The value of this fidelity of the channel estimation is very high in spite of imperfect preparations of test states.
To be specific, the test states that are not (``ideal'') 
pure  states but rather statistical mixtures that are prepared with the fidelity
approximately $0.91$ compared to the ideal test states. 
Simultaneously, we stress that this imperfect test states are completely 
known which is a necessary condition for reliable reconstruction of the 
channel.

The engineered phase damping quantum channels are of particular relevance for
quantum information processing. They represent the most
destructive type of decoherence, because they are destroying 
superpositions of logical qubit states that are necessary for the
success of quantum computing. Therefore, the controlled
implementation of phase damping channels is of use for  testing
of  robustness of quantum computation schemes and
error-correction codes.

\begin{acknowledgments}
We acknowledge financial support via the European Union projects QAP
2004-IST-FETPI-15848, HIP FP7-ICT-2007-C-221889, by the projects
APVV-0673-07 QIAM, GA\v CR GA201/07/0603, VEGA-2/0092/09, 
OP CE QUTE ITMS NFP 262401022, and CE-SAS  QUTE, by the Deutsche 
Forschungsgemeinschaft, and by secunet AG.

\end{acknowledgments}

%%%%%%%%%%%%%%%%%%%%%%%%%%%%%%%%%%%%%%%%%% references
%%%%%%%%%%%%%%%%%%%%%%%%%%%%%%%%%%%%%%%%%%%%%%%%%%%%%%%%%%%%%%%%%%%%


\begin{thebibliography}{10}

\bibitem{nielsen} M.A.~Nielsen and I.L.~Chuang,
{\it Quantum Information and Quantum Computation},
(University Press Cambridge, 2000).

\bibitem{paris}
M.G.A.~Paris and J.~\v Reh\'a\v cek,
{\it Quantum State Estimation},
Springer Series on Lecture Notes in Physics vol. {\bf 649}, (Springer-Verlag, Berlin, 2004).

\bibitem{fisher}
R.A.~Fisher,
Proc. Cambridge Phil. Soc. {\bf 22}, 700 (1925).

\bibitem{fiurasek}
J.~Fiur\'a\v sek and Z.~Hradil,
Phys. Rev. A {\bf 63}, 020101(R) (2001).

\bibitem{sacchi}
M.F.~Sacchi,
Phys. Rev. A {\bf 63}, 054104 (2001).

\bibitem{ruskai}
M.B.~Ruskai, S.~Szarek, and E.~Werner,
Lin. Alg. Appl. {\bf 347}, 159 (2002).

\bibitem{chuang} I.L.~Chuang and M.A.~Nielsen,
J. Mod. Opt. {\bf 44}, 2455 (1997).

\bibitem{poyatos} J.F.~Poyatos, J.I.~Cirac, and P.~Zoller,
Phys. Rev. Lett. {\bf 78}, 390 (1997).

\bibitem{dariano} G.M.~D'Ariano and P.~Lo Presti,
Phys. Rev. Lett. {\bf 86}, 4195 (2001).

\bibitem{jezek} M.~Je\v zek, J.~Fiur\'a\v sek, and Z.~Hradil,
Phys. Rev. A {\bf 68}, 012305 (2003).

\bibitem{altepeter}
J.B.~Altepeter, D.~Branning, E.~Jeffrey, T.C.~Wei, P.G.~Kwiat, R.T.~Thew, J.L.~O'Brien, M.A.~Nielsen, and A.G.~White,
Phys. Rev. Lett. {\bf 90}, 193601 (2003).

\bibitem{Huesmann99} R.~Huesmann, Ch.~Balzer, Ph.~Courteille, W.~Neuhauser, and P.E.~Toschek,
Phys. Rev. Lett. {\bf 82}, 1611 (1999).

\bibitem{Hannemann02}
T.~Hannemann, D.~Reiss, C.~Balzer, W.~Neuhauser, P.E.~Toschek, and C.~Wunderlich,
Phys. Rev. A {\bf 65}, 050303 (2002).

\bibitem{Wunderlich03}
C.~Wunderlich and C.~Balzer,
Adv. At. Mol. Opt. Phys. {\bf 49}, 293 (2003).

\bibitem{Balzer06}
C.~Balzer, A.~Braun, T.~Hannemann, C.~Paape, M.~Ettler, W.~Neuhauser, and C.~Wunderlich,
Phys. Rev. A {\bf 73}, 041407 (R) (2006).

\bibitem{Bell91}
A.S.~Bell, P.~Gill, H.A.~Klein, A.P.~Levick, C.~Tamm, and D.~Schnier,
Phys. Rev. A {\bf 44}, R20 (1991).

\bibitem{Breit31}
G.~Breit and I.I.~Rabi,
Physical Review {\bf 38}, 2082 (1931).

\bibitem{Corney78}
A.Corney,
{\it Atomic and Laser Spectroscopy},
(Clarendon Press, Oxford, 1978).

\bibitem{Brouard03}
S.~Brouard and J.~Plata,
Phys. Rev. A {\bf 68}, 012311 (2003).

\bibitem{Rabenstein04}
K.~Rabenstein, V.A.~Sverdlov, and D.V.~Averin,
JETP Lett. {\bf 79}, 646 (2004).

\end{thebibliography}
\end{document}